\documentclass[prb,amsmath,amssymb,showpacs,1superscriptaddress,floatfix,twocolumn]{revtex4}
\usepackage{graphicx,epsfig}

\begin{document}

\title{Spin current noise and  Bell inequalities in a realistic
superconductor-quantum dot entangler}

\author{Olivier Sauret$^{a,*}$}
\author{Thierry Martin$^b$}
\author{Denis Feinberg$^a$}

\affiliation{
$^a$ Laboratoire d'Etudes des Propri\'et\'es Electroniques des
Solides, CNRS, BP 166,38042
Grenoble Cedex 9, France\\
$^b$Centre de Physique Th\'eorique, Universit\'e de la
M\'editerran\'ee, Case 907, 13288 Marseille, France}

\begin{abstract}
Charge and spin current correlations are analyzed in a source of spin-entangled
electrons built from a superconductor and
two quantum dots in parallel. In addition to the ideal (crossed Andreev) channel, parasitic channels
(direct Andreev and cotunneling) and spin flip
processes are fully described
in a density matrix framework. The way they reduce both the
efficiency and the fidelity of the entangler is
quantitatively described by analyzing the zero-frequency noise correlations of 
charge current as well as spin current in the two output branches. 
Spin current noise is characterized by a spin Fano factor, equal to $0$
(total current noise) and $-1$ (crossed correlations) for an ideal entangler.
The violation of the Bell inequalities, as a test of non-locality
(entanglement) of split pairs, is formulated in terms of the
correlations of electron charge and
spin numbers counted in a specific time window $\tau$. The efficiency
of the test is analyzed,
comparing $\tau$ to the various time scales in the entangler operation.

\end{abstract}

\pacs{73.23.Hk, 74.78.Na, 03.65.Ud}

\maketitle

\section{Introduction}

The production and the analysis of entangled states  
in condensed matter devices have
recently emerged as a mainstream in nanoelectronics.
Indeed, entanglement between electrons, besides checking fundamental quantum  properties
such as non-locality, could be used for
building logical gates and quantum communication devices \cite{IQ}.
One may choose to consider nanoscopic devices where
electrons behave essentially as free particles \cite{free}, in a way similar to photons in quantum optics.
Alternatively, exploiting electron interactions open new possibilities. 
This allows for instance to use the
electron spin as a qubit in quantum dots \cite{loss_divincenzo}, owing to spin
relaxation/coherence times ranging in semiconductors
from fractions of $\mu s$ (bulk \cite{coherence_bulk}) to fractions of $ms$
($T_1$ in quantum dots \cite{coherence_dot}).
Spin entanglement \cite{lesovik_martin_blatter,loss,saraga}
and teleportation \cite{TP} scenarios have been proposed within such a framework.
A test of this spin entanglement was previously proposed with the help of 
a Bell diagnosis \cite{kawabata} , 
for a normal metal-superconducting device described within the context of 
scattering theory, operating in the ideal crossed Andreev regime
\cite{chtchelkatchev}. 
However, in any device several unwanted electronic
transitions may spoil entanglement and introduce decoherence. It is of key importance
to check the device robustness against these parasitic effects and to show how to diagnose them through transport
properties. Owing to
the strategic role of entangled pairs in quantum information, we focus here on
the source of spin-entangled electrons proposed
by Recher et al. \cite{loss}. Following a detailed study of the average current
due to the various processes \cite{sauret_bloch}, the present work addresses noise correlations
as a diagnosis of the device operation.

\begin{figure}[h]
\epsfxsize 8.6cm
\centerline{\epsffile{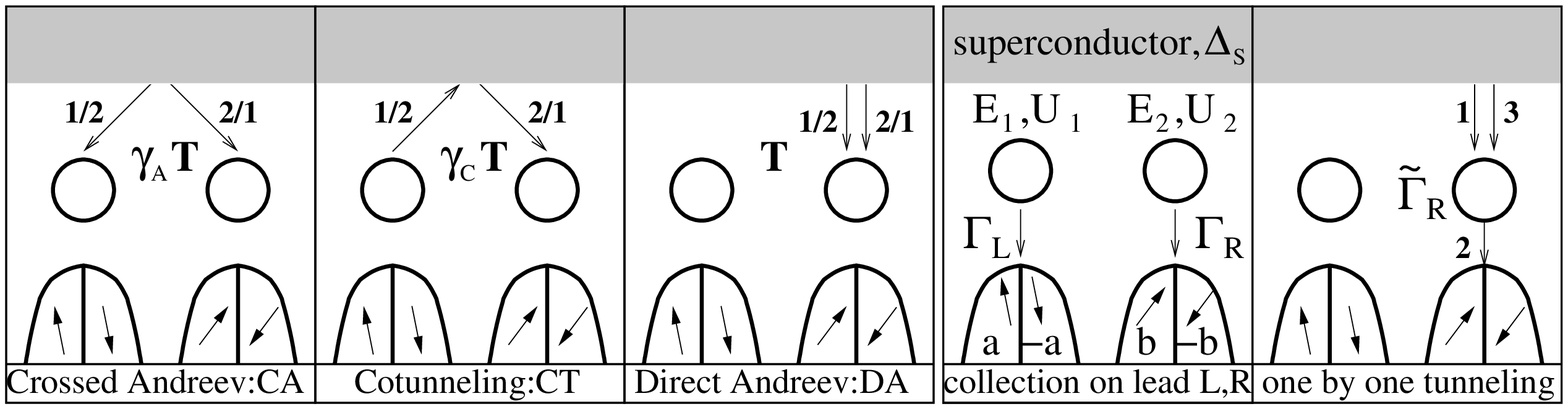}}
\caption{Processes involved in the operation of the entangler. Pairs are emitted by the
superconductor (gray) towards the two quantum dots, electrons are then collected by the leads
$L,R$ along predefined spin polarizations
$\pm a$, $\pm b$ ($4^{th}$ panel). The
order of elementary transitions involving
virtual states is indicated :
Crossed Andreev (CA) produce split pairs and triggers entanglement.
CT (cotunneling),
Direct Andreev (DA) as well as
one-by one tunneling through the same dot spoil entanglement. The three left panels describe coherent
transitions, while the two right ones describe incoherent transitions.}
\label{fig1}
\end{figure}

The entangler of Ref. \onlinecite{loss}
is made from a superconductor (S), adjacent to two small quantum dots in parallel, each being
connected to a normal lead featuring a quantum channel. The dots are assumed to filter electrons one by one in
a single orbital level, and
Cooper pairs emitted from the superconductor are split
in the two dots -- the so-called crossed Andreev channel (CA) \cite{CA,choi_bruder_loss}. Ideally,
the constituent electrons
then propagate in the two output leads as an entangled pair.
Parasitic quantum processes may also occur in the entangler, they include elastic cotunneling
(CT) \cite{CA,CT} between the two dots via S,
and tunneling of a pair through
the same lead by a Direct Andreev process (DA) (Fig. \ref{fig1}). The main (CA) channel as well as DA
were studied in Ref. \onlinecite{loss}.
Using perturbation theory, the average current contributions to CA and DA
were separately calculated.
In the same spirit, Ref. \onlinecite{buttiker} used a
beam splitter on the entangler's output, in order to detect the
entangled singlets by their noise
bunching properties \cite{burkard}. In reality, all these transport
channels are mixed together.
Instead of analyzing them separately, the full electron flow should be
studied self-consistently, and be eventually characterized by its density matrix.
The present work is based on a quantum master
equation scheme which treats all processes on an equal footing
\cite{sauret_bloch} and gives access to the full
density matrix of emitted electrons. It is used
to compute the charge and spin correlations in the current flowing
through the two branches
of the entangler. In particular,
spin current correlations  \cite{sauret_feinberg}, even taken at zero
frequency, are shown to quantify the
entangler's efficiency and fidelity.
Next, the violation of Bell inequalities (BI) can be tested
by computing the cross correlators of charge and spin particle
numbers in a given time window.

\section{Ideal regime and the other transport processes}

Transport is described by the
processes connecting the different states --
identified by the charge number and spin of the dots. Extra electrons $n_1$, $n_2$
define the charge states ($n_1n_2$) in the dots. The electronic processes at work in the device include
both coherent processes (CA, CT and DA) and incoherent processes
(transitions between the dots and the leads), see Fig. \ref{fig1}.
In the ideal operation of this entangler,
the Coulomb blockade prevents double occupancy in each dot, so that DA is forbidden. Also, the
superconducting gap is supposed to be very large, so that one-by-one tunneling
(last panel in Fig. \ref{fig1}) is also excluded. In addition, CT is neglected
in this ideal regime, owing to the energy difference between the electron states on the two dots.
Then, starting from an empty orbital state (denoted ($00$)), crossed Andreev (CA)
reflection couples to
a singlet state shared between the two dots, ($11_s$)
with an amplitude $\gamma_A T$ ($\gamma_A \ll 1$ is the geometrical
factor \cite{loss,CA,choi_bruder_loss}).
For a resonant CA transition, the dot energy levels $E_i$ satisfy $E_1+E_2=0$.
The two electrons are collected into the leads
(with chemical potentials $\mu_{L,R}<E_1,E_2$) by transitions to
states ($01$) or ($10$), then ($00$),
with rates $\Gamma_i$ ($i=1,2$).

Let us now consider parasitic transport processes.
If the intradot Coulomb
charging energy $U$ is not very strong, a
coherent transition from ($00$) to a doubly
occupied dot state ($20$) or ($02$) can occur via a direct Andreev (DA)
process \cite{loss}. Notice that this occurs with an amplitude $T$ larger than for CA, since
it is not affected by the geometrical reduction factor $\gamma_A$.
Electrons are successively collected
by a single lead, with rates $\Gamma'_i$ then $\Gamma_i$, eventually
reaching the empty state ($00$).
This transport channel
implies dot energies $2E_i+U$. In addition, DA can also
start from an initial state ($10$) or ($01$), proceeding through states
($21$) (respectively ($12$)) with energies
$2E_1+E_2+ U$ (respectively $E_1+2E_2+ U$).
Further collection into the leads either give states ($20$), ($02$), or
singlet and triplet
states ($11_{s,t}$). The mixing of triplet states with the
desired singlets may cause decoherence in the source operation.
In another process, the two electrons from a Cooper pair can tunnel
one by one towards the same lead \cite{loss}, involving a singly occupied virtual
state which costs
an energy at least equal to the superconducting gap $\Delta_S$ (Fig.
\ref{fig1})).
Contrary to the DA process, the dot is emptied before the
quasiparticle in S is annihilated. The state ($10$) (or ($01$)) is reached by
an incoherent process with a rate $\tilde{\Gamma}_{i} \sim \Gamma_i
T_i^2/\Delta_S^2$.
Last, CT involves a coherent transition of an electron from one dot
to the other
via S, and mixes all the above processes \cite{sauret_bloch}: it couples state ($01$) to
($10$), but also ($20$) (or ($02$)) to ($11$),
($21$) to ($12$). CT has an amplitude $\gamma_C T \ll T$, of the same order of magnitude as
that of CA.

The density matrix equations involving populations (diagonal
elements) and coherences
(nondiagonal elements) include these processes
altogether in a consistent and non-perturbative
manner. Here one assumes weak couplings $\Gamma_{L,R} < k_B\Theta$ of the dots to the output
leads $L$, $R$ ($\Theta$ is the temperature). On the other hand, the leads are biased such as
$|\mu_{L,R}| > T$, $\Gamma_{L,R}$, ensuring irreversible collection of the electrons in $L$, $R$.
The density matrix equations have been derived and detailed in Refs. \onlinecite{sauret_bloch,these}, where
the average currents in each branch have been computed. From these Bloch-like equations,
one can also calculate the conditional
probabilities -- for a transition to a given state,
assuming a previous initial state -- which enter the
calculations of the noise correlators \cite{korotkov,note,BB}. Processes involving 
states with at most two electrons on the double dots are represented on Fig. 6 of 
Ref. \onlinecite{sauret_bloch}, yet three-electron states are also 
included in the calculations.

\section{Charge and spin current correlations}

Let us assume that the leads contain separate channels perfectly filtering spin currents
polarized along two predefined and opposite directions.
An understanding of the charge and spin correlations
can be obtained
by considering the spin-resolved zero-frequency noise \cite{sauret_feinberg} :
$I_{\sigma}$ being the current carried by electrons with spin $\sigma=\pm a$
 (along a given spin direction $\bf a$), we define 
\begin{equation}
S_{ij}^{\sigma\sigma'}(0)=
\int dt \langle \{\Delta I_i^{\sigma}(t),\Delta I_j^{\sigma'}(0)\}\rangle
~,
\end{equation}
where $i,j=\{L,R\}$ and $\Delta I_i^{\sigma}(t)$ is the deviation
from the average current component $\langle I_i^\sigma (t)\rangle=I_i^\sigma$.
We assume that the lead resistances and couplings to the dots are not
spin-dependent. Because 
the superconductor is emitting singlet pairs, it results that
 the average current in each lead is not spin-polarized,
e.g. $I_i^\sigma=I_i/2$. By definition, the charge and spin
current noise correlation read:
\begin{eqnarray}
S_{ij}^{ch} &=& \sum_\sigma (S_{ij}^{\sigma \sigma} +
S_{ij}^{\sigma -\sigma})
\\
S_{ij}^{sp} &=& \sum_\sigma (S_{ij}^{\sigma \sigma} -
S_{ij}^{\sigma -\sigma})~.
\end{eqnarray}
The latter simply expresses the time correlations of the spin current
$I_i^{spin} = I_i^{\sigma}-I_i^{-\sigma}$. The average total charge current passing through the entangler
can be written as 
\begin{equation}
I=I_L+I_R,~
\end{equation}
where the currents
in leads $L$ and $R$ can be separated in two components, as
\begin{eqnarray}
I_L &=& 2I_{LL}+I_{LR}~, \\
I_R &=& 2I_{RR}+I_{LR}~.
\end{eqnarray}
Here $I_{LL}$, $I_{RR}$ and $I_{LR}$
respectively count pairs passing in unit time through $L$, or $R$, or
through $(L,R)$ as split (entangled) pairs. These current components can be written as:
\begin{eqnarray}
I_{LR}&=&\frac{1}{2}p_SI~, \\
I_{LL}&=&\frac{1}{2}p_LI~, \\ 
I_{RR}&=&\frac{1}{2}p_RI~,
\end{eqnarray}
with probabilities such that $p_L+p_R+p_S=1$.

It is crucial to notice here that all the electrons passing through the
entangler enter the two-dot system by pairs, emitted in an Andreev process (CA, DA or one-by-one tunneling).
Therefore, the spin current fluctuations
due to electrons of a given pair are correlated, since they track the passage of
singlets. On the contrary, electrons emitted within different pairs display no
spin correlation whatsoever. It results that, while calculating the correlations
of spin currents at different times,
the contributions of separate pairs drop out, so that spin noise in the leads
uniquely tracks the spin correlations inside a pair. These correlations should be non-ideal
since they track what
 remains from the emitted singlets
after their passage through the two-dot system. This property of probing single pairs 
strongly contrasts
with the charge current fluctuations which - in a sequential tunneling  process -  
correlate successive pairs due
to the Pauli principle and the Coulomb interaction in the dots. Therefore the spin current noise
is very well suited to study the spin correlations inside pairs with zero-frequency noise.
To illustrate this, it is convenient to write the various spin noise correlations as follows :
\begin{eqnarray}
\label{eq:spin noise}
\nonumber
&S_{LL}^{sp}=2eI_L-4eI_{LL}(1-f_L)=2eI[p_Lf_L+\frac{1}{2}p_S]~,\\
\nonumber
&S_{RR}^{sp}=2eI_R-4eI_{RR}(1-f_R)=2eI[p_Rf_R+\frac{1}{2}p_S]~,\\
&S_{LR}^{sp}=S_{RL}^{sp}=-2eI_{LR}(1-f_S)=-eIp_S(1-f_S)~.\nonumber\\
\end{eqnarray}
Let us comment the various terms in Eq. (10).
The first term in $S_{ii}^{sp}$ corresponds to the autocorrelation of
electron wave packets \cite{korotkov,sauret_feinberg}, and takes a Poisson value. The other terms
are negative and reflect the spin anticorrelation inside singlets.
The Fano-like reduction factors $f_i$, $f_S$ quantify this correlation. For instance,
blocking one lead ($\Gamma_R=0$ for instance), all pairs pass through
$L$ ($p_L=1$) in a sequential way.
In this case $I=I_{L}=2I_{LL}$. Moreover, since the dot filters electrons one by one, 
one easily checks that $f_L=0$, thus the spin noise through $L$ is
zero. This "spin Fano factor" equal to zero is
 a property of a S/N junction, characteristic of an unspoiled Andreev process 
\cite{sauret_feinberg} : at a superconductor/normal (S/N)
metal junction, zero spin noise is the fingerprint of charge carriers being paired in
a singlet state.
On the other hand, if the entangler operates ideally (only CA), then
$p_S=1$, $f_S=0$ and $S_{ii}^{sp}=eI$. This is somewhat similar to the behaviour 
of a quantum dot in the sequential regime, with a single
orbital state involved in transport and strong Coulomb charging energy, 
yielding a spin Fano factor equal to one \cite{sauret_feinberg}. 

On the other hand, the crossed spin correlation is $S_{LR}^{sp}=-eI$. 
This value corresponds to an effective spin Fano factor
$-1$, and serves as a reference value for an ideal
operation of the device. These Poisson-like Fano factors, confirmed -- see below -- by 
the solution of the master 
equations, are valid for any value of the couplings of the dots to the leads. This is in contrast 
with the "charge" Fano factors in resonant quantum dot devices which reach the value one only in the 
very asymmetric limit, where correlations between successive charges crossing the dots 
become negligible. 

Summing all contributions in Eq. (10), the spin noise of the total current is
\begin{equation}
S_{tot}^{sp}=S_{LL}^{sp} + S_{RR}^{sp} + S_{LR}^{sp} + S_{RL}^{sp} = 2eIF_S~,
\end{equation} 
with the total spin Fano factor
\begin{equation}
F_S=p_Lf_L+p_Rf_R+p_Sf_S~.
\end{equation} 
Notice that in our definition the spin Fano factors measure the zero-frequency
spin current noise in units of the average {\it charge} current 
(not the average spin current which is zero).

In the general case, when all (CA, DA and CT) processes come into play, pairs are both
distributed (e. g. they exit through $LL$, $RR$ and $LR$)
{\it and} "mixed" together (electrons of a pair may intercalate between electrons of another)
 : for instance, due to states with three electrons
in the double dot, two electrons of a split pair $L,R$ can be
separated by a time interval during which one or more
pairs pass through a single dot (as illustrated by the sequence ($00$),
($11$), ($10$), ($12$), ($11$), ($10$), ($00$)).
This variable delay between two spin-correlated electrons is
responsible for the factors $f_{L,R,S} > 0$ which measure the degree of mixing.

This preliminary analysis suggests that parasitic processes have two
effects. First, they reduce the probability of split (entangled) pairs over non-entangled ones,
thus $p_s$ defines the entangler {\it efficiency}. And secondly, they reduce the spin
correlation of a split pair, $1-f_s$ defining the entangler
{\it fidelity} with respect to the singlet state. Notice that both efficiency and fidelity
are defined here with respect to a given
 measurement probe, e. g. zero-frequency noise, or alternatively 
time-resolved correlation in the present work.

\section{Quantitative analysis}

We now study these trends
quantitatively. Charge and spin current noise correlations are calculated
from the entangler quantum master equations, derived in ref.  \onlinecite{sauret_bloch},
where four-electron states in the dot pair are neglected for simplicity. 
We successively focus on the
correlations of the total current $I_L+I_R$, of the separate currents $I_L$, $I_R$, 
and on the crossed correlations between $L$ and $R$.

\begin{figure}[h]
\epsfxsize 8.5cm
\centerline{\epsffile{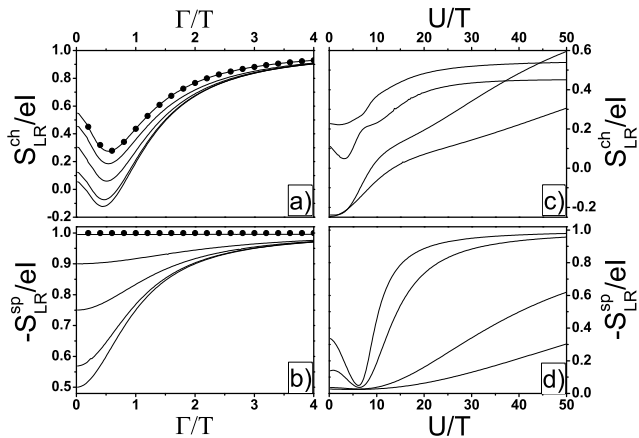}}
\caption{Spin current cross-correlations, assuming CA to be resonant:
$E_1+E_2=0$, and $\Gamma_L=\Gamma_R$.
$\Delta_S$ is the
largest energy scale, thus one-by-one tunneling is negligible. $T=1$ and 
$\gamma_A=0.2$ is assumed. 
In $a)$ and $b)$, only CA and CT are taken into account,
assuming $U\rightarrow\infty$.
Charge and spin cross-correlations are
plotted for $0<\Gamma<2$ and several values of $E=|E_1-E_2|=
\{0,0.2,0.5,1,5 \}$ (bottom to top).
The dotted line features the ideal case.
$c)$ and $d)$ correspond to $0<U<50$ and $E=5$
(CA and DA, but no CT), for $\Gamma_{L,R} = \{0.01,1,5,10\}$
(top to bottom).}
\label{Sspin}
\end{figure}

\subsection{Ideal regime}
Consider the ideal operation (CA only) for different
values of $\Gamma$ (assuming the Andreev amplitude to be equal to $T=1$, and a geometrical
factor $\gamma_A = 0.2$). All electron pairs
are split and cross the device sequentially (one after the other).
The average (charge) current\cite{sauret_bloch} $I$ (see Fig. 4a)
displays a maximum in the region
$\Gamma \sim \gamma_AT$. This maximum can be easily explained by the balance between the entrance
and exit rates of the two electrons in the double dot. The total charge noise 
through the entangler (see Fig. 5b, and also Ref. \onlinecite{these}):
\begin{equation}
S_{tot}^{ch}=S_{LL}^{ch} + S_{RR}^{ch} + S_{LR}^{ch} + S_{RL}^{ch}
\end{equation}
displays a minimum in the same range of parameters, as expected. Moreover, for large $\Gamma/T$, it reaches the value
$4eI$. This doubling of the Fano factor with respect to the Poisson value signals the
passage of electron pairs \cite{doubling_noise}.
On the other hand, the crossed charge noise correlation $S_{LR}^{ch}$ is positive (Fig.
\ref{Sspin}a), in contrast to the partition noise of single electrons, which is negative \cite{BB}.
This reflects the perfect charge correlation of split pairs
\cite{positive_correlations}.
As a function of $\Gamma/T$, $S_{LR}^{ch}$ also displays a minimum (Fig.
\ref{Sspin}a) and becomes Poissonian
in the limit $\Gamma \gg \gamma_AT$, where successive pairs are collected much faster than emitted,
so are well-separated in time. Notice that in the ideal operation
one has:
\begin{equation}
S_{LR}^{ch}=S_{LL}^{ch}=S_{RR}^{ch}~.
\end{equation}

Turning to the spin noise, one finds that the total spin noise $S_{tot}^{sp}$ is zero, 
as discussed above. Moreover, one
verifies that: 
\begin{equation}
S_{LL}^{sp}=S_{RR}^{sp}=-S_{LR}^{sp}=eI~,
\end{equation}
thus diagnosing perfect spin correlations
from the entangler.

\subsection{General regime}
Parasitic processes are next illustrated in presence of
cotunneling alone ($U$ is chosen as infinite), which introduces a coupling between
branches $L$, $R$.
In this case, pairs are distributed ($LL$, $RR$ and $LR$) but still sequentially
(they pass one after the other and do not mix) : therefore all the $f$'s in Eq. (1) are equal to zero.
First, assuming $\Gamma_L = \Gamma_R$, the average current as well as
$S_{tot}^{ch}$ which sums over all branches are not modified \cite{footnote2}. On the contrary,
$S_{LR}^{ch}$ decreases (Fig. \ref{Sspin}a)
and can even become negative when cotunneling has a strong effect : due
to branch coupling, charge cross-correlations are lost and partition noise may dominate over
pair correlations. On the other hand, spin noise gives:
\begin{equation}
S_{ii}^{sp}=-S_{LR}^{sp} = eIp_s~~,
\end{equation} 
Therefore the spin Fano factor simply counts the split pairs
(Fig. \ref{Sspin}b) and measures the entangler efficiency.
One checks that the total spin noise satisfies:
\begin{equation}
S_{LL}^{sp}+ S_{RR}^{sp} +
2S_{LR}^{sp}=0~,
\end{equation}
i. e. the whole entangler is a perfect Andreev source :
whatever the origin of the pairs ($LL$, $RR$ or $LR$), they are well-separated and the spin
correlations of successive electrons is perfect.
Notice that $p_s$ tends to one for large $\Gamma \gg
\gamma_CT$ where CT has no time to occur and becomes irrelevant.
In practice, for large $|E_1-E_2|/T > 5$,
CT is merely suppressed. In summary, with CT alone we have a situation where the
efficiency of the entangler is reduced (and measured by the spin noise),
while its fidelity remains perfect. Yet, notice that
further use of these pairs in the output quantum channels 
would require an additional device to collect split pairs $LR$
and filter out the others.

Next, double occupancy in the dots is allowed (process DA). Fixing $\Gamma/T$, the
crossed charge noise is reduced and
can become negative at small $U$ (Fig. \ref{Sspin}c). Spin
noise correlations
also decrease as $U/T$ decreases (Fig. \ref{Sspin}d). The structures observed at
small $U$ for small $\Gamma$ are due to the difference between the single electron levels
$E_1$, $E_2$, and DA transitions becoming resonant when $E_i \sim -U$.
The width of the minimum increases for large $\Gamma$,
due to two competing processes.
On one hand, the probability for having a split pair (CA)
in the 2 dots oscillates on a large time scale $(\gamma_AT)^{-1}$. On the other hand,
the probability for a DA processes is smaller, but it oscillates
more rapidly\cite{sauret_bloch} ($(U^2+4T^2)^{-1/2}$). A high detection rate thus
dynamically favors the fast DA process, even though it is weaker. In the
resonant regime $E_i \sim -U$, $S_{LL}^{ch}$ can even approach $4eI_L$, e.g.
 noise doubling in branch $L$. Moreover, it is important to point out
that due to direct Andreev processes, the total spin noise is no more equal to zero. Actuelly, 
the occurrence
of three-electron states in the double dot has the effect of mixing the pairs, reducing
the fidelity, especially for small $U$ and $\Gamma$. 

In summary, the analysis of
this quite general case demonstrates that
optimal pair correlations are obtained if $\gamma_AT < \Gamma < U$. Nevertheless, at fixed
$\Gamma/U$,  a too small
$\gamma_AT/\Gamma$ ratio is detrimental, as it favours $LL$ and $RR$ pairs by a
dynamical effect.

\section{Bell inequality test}

In quantum optics, entanglement is typically probed by a Bell inequality (BI)
test. It relies on the
measurement\cite{aspect} of number correlators.
One may simplify the response of the electronic circuit such as to measure,
not the instantaneous current, but instead the particle number
accumulated during a time $\tau$ is: 
\begin{equation}
N_\alpha(t,\tau) =\int_t^{t+\tau} dt^\prime\,
I_\alpha (t^\prime)~,
\end{equation}
($\alpha=\pm a, \pm b$ 
see Fig. \ref{fig1}).
The inequality which is derived assuming a product density matrix
weighted by local hidden variable reads\cite{chtchelkatchev}:
\begin{subequations}
    \begin{eqnarray}
    \label{Bell_eq}
    |G({\bf a,b})-G({\bf a,b^\prime})+ G({\bf a^\prime,b})
    +G({\bf a^\prime,b^\prime})|\leq 2,
\\
    G({\bf a,b})
    =\frac{\langle(N_{a}(\tau)-N_{-a}(\tau))(N_{b}(\tau)
    -N_{-b}(\tau))\rangle}{\langle(N_{a}(\tau)+
    N_{-a}(\tau))(N_{a}(\tau)+N_{-a}(\tau))\rangle}
\end{eqnarray}
\end{subequations}
In Ref.\onlinecite{chtchelkatchev}, the correlators
$\langle N_{\alpha}(\tau)N_{\beta}(\tau)\rangle$ were related to
zero-frequency current noise
correlators via an approximate relation. 
This approximation breaks down for short 
times. 
We follow
Ref.\onlinecite{lebedev} and calculate the
correlators from first principles.
The cross correlator for arbitrary spin directions of the filters can
be separated into a parallel
and an antiparallel component:
\begin{eqnarray}
\label{S_ideal_long}
\langle N_{\alpha}(\tau)N_{\beta}(\tau)\rangle &=&
\langle N_{L\uparrow}(\tau)N_{R\downarrow}(\tau)\rangle
\sin^2({\theta_{\alpha\beta}}
/2)
\nonumber\\
&~&
+\langle N_{L\uparrow}(\tau)N_{R\uparrow}(\tau)\rangle
\cos^2({\theta_{\alpha\beta}}/2)~,
\end{eqnarray}
with $\theta_{\alpha\beta}$ the relative angle between the
polarizations in $L$ and
$R$. With the same choice of angles as considered in Ref.
\onlinecite{chtchelkatchev} the BI becomes:
\begin{equation}
\label{Bell_maximum}
|\Delta N^{sp}_{LR}/(\Delta N^{ch}_{LR}+\Lambda^+)|
\leq 1/\sqrt{2}
\end{equation}
with 
\begin{equation}
\Delta N^{sp}_{LR}=\langle
N_{L\uparrow}(\tau)N_{R\downarrow}(\tau)\rangle
- \langle N_{L\uparrow}(\tau)N_{R\uparrow}(\tau)\rangle~,
\end{equation}
 and 
\begin{equation} 
 \Delta
N^{ch}_{LR}+\Lambda^+
=\langle N_{L\uparrow}(\tau)N_{R\downarrow}(\tau)\rangle  +
\langle N_{L\uparrow}(\tau)N_{R\uparrow}(\tau)\rangle~,
\end{equation}
where 
\begin{equation}
\Lambda^+=\tau^2\langle I_L\rangle\langle I_R\rangle
\end{equation} 
is the reducible part of the charge correlator.

\begin{figure}[h]
\epsfxsize 8.5 cm
\centerline{\epsffile{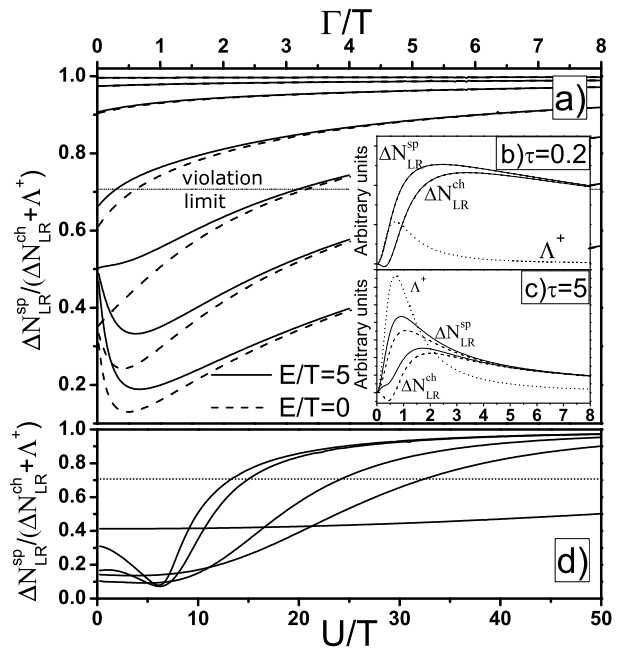}}
\caption{The left part of the BI is compared to the
limit $2^{-1/2}$ (one assumes $T=1$ and 
$\gamma_A=0.2$). $a)$ It is plotted
for $0 < \Gamma_{L,R}< 8$, two values of $E$ ($0$, and $5$ which
simulates the ideal working regime),
and $\tau T/h=\{0.2,0.5,1,5,10,20\}$ (from the top to the bottom). To
characterize the BI test, each contribution of $|\Delta N^{sp}_{LR}|/(\Delta
N^{ch}_{LR}+\Lambda^+)$ is plotted for
$\tau=0.2$ b), $\tau=5$ c).
d) Double occupancy is taken into account, assuming
$E=5$, $\Gamma_{L,R}=\{0.01,1,5,10,100\}$ (top to bottom) and $\tau=0.2$}
\label{Bell}
\end{figure}

In the ideal case (see Fig. 3a),
the BI violation is maximal for $\tau < h/T$ ($h$ is the Planck constant). Increasing $\tau$, thus
the number of measured pairs,
quantum spin correlations show a minimum at $\Gamma \sim \gamma_AT$,
due to a decrease of
$\Delta N_{12}^{sp/ch}/\Lambda^{+}$ (Fig. 3b,c). BI violation is
recovered provided
$\Delta N_{12}^{ch}  \sim \tau I$ dominates over $\Lambda^{+}=\tau^2
I^2$, where
$I \sim \gamma_A^2 T^2/\Gamma$ for large $\Gamma$ and $I \sim \Gamma$
for small $\Gamma$. This means
that pairs should be measured roughly one by one
\cite{chtchelkatchev}. Now, considering
parasitic processes, CT tends to prevent BI violation at small $\Gamma$,
and for large enough $\tau$ (Fig. 3a).
Finite U sensibly decreases the quantum spin
correlations (Fig. 3d). The minimum flattens as $\Gamma$ increases, as in Fig. 2d (dynamical effect). Yet,
comparing Figs. 2 and 3, one sees that the BI test is much less
affected by parasitic processes than the spin noise. For instance, with $\Gamma = 5$ and $U/T =
40$, the spin noise is about $0.5$, signalling a low efficiency, while the BI is
violated with $\tau = 0.2$.
In fact, this window allows filtering of split CA pairs, mostly
dropping "wrong" DA pairs. Therefore,
tuning $\tau$ to an optimum value yields high fidelity of entanglement,
even if the efficiency of the entangler is low.

\section{Spin-flip processes}

Finally, let us address the effect of spin-flip scattering
in the dots, here simply described by a spin relaxation time $\tau_{sf}$.
As entanglement is concerned, this
acts as a decoherence source, as it induces some mixing of singlet
states ($11_s$) with triplet states ($11_t$).
Spin-flip is easily incorporated into the general density matrix equations \cite{these},
in a way similar to Ref. \onlinecite{sauret_feinberg}. It consists in defining occupation
states ($n_{1\sigma},n_{2\sigma}$), and adding into the density matrix equations
the terms corresponding to the decay of the spin densities $n_{i\sigma}-n_{i-\sigma}$.
The effect of spin relaxation is summarized on Fig. 4. First, surprisingly enough,
the average current is strongly quenched as soon
as the spin-flip rate is of the order of
the CA amplitude (Fig. 4a). This can be explained as follows. CA is coherent provided it
involves a spin-conserving transition between degenerate states.
 Spin-flip in the dot, if faster than the CA
resonance between states ($00$) and ($11_s$), suppresses this
resonance, opening a new decay channel. Indeed, spin flip competes with CA in a
way similar to charge decay toward $L$, $R$. For $\tau_{sf} < h/\Gamma <
h/\gamma_AT$, the current behaves as
$\tau_{sf}^2$. For the same reasons, spin-flip also decreases charge correlations (Fig.
4b). Secondly, as expected,
spin-flip decreases the spin noise correlations in leads $L,R$ (Fig.
4b), by essentially diminishing the fidelity $1-f_S$. In addition, Bell correlations are also
affected by spin-flip, but, as above, much less than spin noise.

\begin{figure}[h]
\epsfxsize 8.5 cm
\centerline{\epsffile{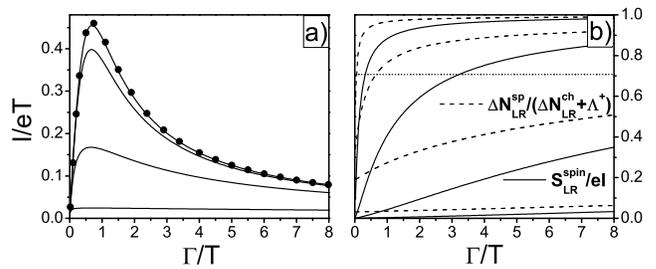}}
\caption{(a) Average current I ; (b) spin current noise (continuous)
and Bell correlation (dotted),
as a function of $\Gamma/T$, with $U=\infty, E=5$ (CA only) and finite spin-flip time $\tau_{sf}$
in the dots. $\tau_{sf} T/h=\{10, 1, 0.1, 0.01\} $ decreases from top to bottom.}
\label{decoherence}
\end{figure}

\section{Discussion}

Let us now discuss the relevance of the above analysis in view of a future experimental
realization of this entangler. First, we summarize the required operation regime, from the above
as well as previous discussion \cite{loss, sauret_bloch} :

\begin{eqnarray}
&&\Delta_S, U, |E_1-E_2| > \delta \varepsilon > \mu_{L,R}, k_B \Theta, \tau_{sf}^{-1} \nonumber
\\
&&~~~> \Gamma_{L,R} > \gamma_AT,\gamma_CT
\end{eqnarray}

\noindent
where $\mu_{L,R}$ is the voltage drop between the superconductor and the leads $L,R$ and
$\delta\varepsilon$ the separation between different orbital states in the dots. We assume that the average
currents in both leads can be measured, as well as zero-frequency (charge and spin) noise correlations.
Niobium will be taken for the superconductor ($\Delta_S \sim 9.2K$), and $\delta\varepsilon \sim 1meV \sim 12K$
for GaAs or InAs-based small quantum dots. A reasonable value for $U$ is $45K$, larger than $\Delta_S$,
which does not change the above conclusions, for which $U < \Delta_S$ is instead assumed. A typical relaxation rate
is $\Gamma_{L,R} \sim 100mK$, larger than $k_B \Theta$. One can then discuss the operation of the entangler as
a function of the CA amplitude, say the parameters $T$ and $\gamma_A$. As an experimental constraint, we fix the
value of the current to a minimum of $1pA$, to make current correlations measurable. 
The current is plotted in Fig. 5a in nanoAmperes as a function of the electronic amplitude T, 
expressed in Kelvins, for two values of $\gamma_A$. 

\begin{figure}[h]
\epsfxsize 8.5 cm
\centerline{\epsffile{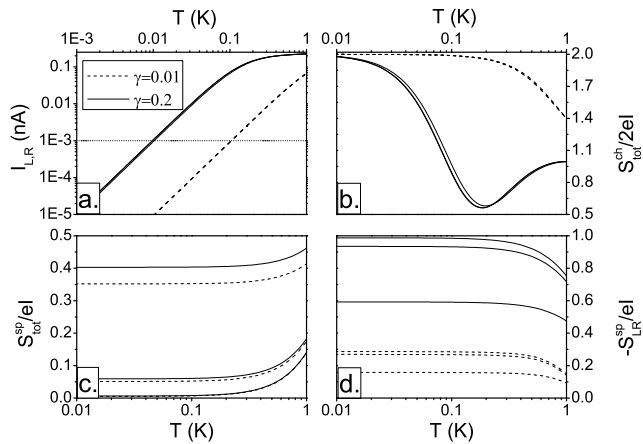}}
\caption{Various diagnosis of the entangler, plotted as function of the pair 
tunnel amplitude $T$ (in Kelvins), for 
$\gamma_A = 0.01$ and $0.2$. a) The average current, assumed larger than 1 pA; b) The total 
charge noise; c) The total spin noise. The spin-flip time decreases from bottom to top, 
$\tau_{sf}(ns) = \{100, 10, 1\}$; d) The crossed spin noise. The spin-flip time decreases from top to bottom, 
$\tau_{sf}(ns) = \{100, 10, 1\}$, causing nearly no variation in panels a,b.}
\label{Cas_reel}
\end{figure} 

For optimum operation, the entangler should provide pairs which are i) well-separated in time;
ii) split as much as possible in ($L,R$); iii) in the singlet state.
As a first probe, the total charge noise tests the temporal separation of successive pairs,
signaled by the doubling of shot noise, $S_{tot}^{ch} = 4eI$ (Poisson result).
We have numerically verified (see Fig. 5b) that this criterion is fulfilled 
up to a few percent if $\gamma_A T \leq \Gamma_{L,R}$. It is compatible with our constraint on $I$ 
for $T \sim 0.1 K$ ($\gamma_A = 0.01$) and $T \sim 0.01 K$ ($\gamma_A = 0.2$). 

As a second probe, the total spin noise
also reflects the "mixing" of pairs in time, being zero if pairs are well-separated ($T \leq \Gamma$
whatever $\gamma_A$ without spin flip). Fig. 5c shows that this criterion is very well fulfilled (lower curve).
In addition, $S_{tot}^{sp}$ is sensitive to spin-flip. Therefore a combination of total charge and spin noise
gives information on pair mixing and spin flip. Spin flip tends to increase $S_{tot}^{sp}$ from the 
ideal value $0$, especially for large $\gamma_A$.

As a third criterion, crossed spin correlations
provide information on the fraction $p_S$ of split pairs ($L,R$). We find that $\gamma_A = 0.2$
gives an excellent ratio (nearly $1$ for $T \leq \Gamma_{L,R}$). On the contrary, for $\gamma_A = 0.01$,
this ratio drops to about $0.2$, $LL$ and $RR$ pairs being "dynamically" favored.
Unless some further processing of pairs is achieved, this is detrimental to the entangler operation. 

To summarize this analysis, a small value of $\gamma_A$ is preferable for pair separation, but 
a large one is requested for the singlet fidelity. Given  these conflicting requirements, 
a good compromise is obtained 
with large $\gamma_A \sim 0.2$ but small $T \sim 0.01 K$. Concerning the spin-flip time, values 
of the order of $10 h/T \sim 50 ns$ are suitable. 

\section{Conclusion}
We have analyzed in detail the operation of a realistic superconductor-dot entangler. 
We have shown that zero frequency charge and spin current correlations
allow a detailed analysis of the efficiency and fidelity of
the entangler in terms of parasitic and spin flip processes. Spin Fano factors for spin 
current correlations appear as an optimum characterization of the entangler operation. Due 
to the absence of spin correlation of electrons emitted within different Andreev 
processes, such spin Fano factors directly probe the spin correlations within an entangled pair. 
Depending on the quantity under interest, they are equal to $\pm 1$ in the ideal operation regime. 
In this sense, measuring the spin current noise is an alternative to time-resolved measurements 
which should capture pairs one by one. Yet, for an absolute diagnosis of entanglement, a time-resolved
Bell inequality measurement is necessary, within a narrow enough time window. Contrarily to low 
frequency measurements, this allows to check the
 entangler fidelity even in presence of a low efficiency.
As a result of this work, we conclude that a satisfactory operation is
upon reach with realistic parameters.
For detection rates $ \Gamma\sim 100mK$, the shortest
time window described
for a Bell test corresponds to $\tau\sim 1 ns$, a time scale which could be
accessible with fast electronics apparatus. Moreover, the geometric parameter
$\gamma_A$ appears to be crucial and should be rather large. Encouraging results
have recently been obtained in two experiments which have observed crossed Andreev reflections at
distances not very small compared to the superconductor coherence length \cite{exptCA}.

LEPES is under convention with Universit\'e Joseph
Fourier. Further support from Ministry of Research
``A.C. Nanosciences NR0114'' is gratefully acknowledged.

*  Present address : Department of Theoretical Physics,
Budapest University of Technology and Economics, Budapest, Hungary.


\begin{thebibliography}{99}
%
%
\bibitem{IQ} M. A. Nielsen and I. L. Chuang, Quantum Computation and Quantum Information
(Cambridge University press, 2000).
%
\bibitem{free} C. W. J. Beenakker, C. Emary, M. Kindermann and J. L.
van Velsen, Phys. Rev. Lett {\bf 91}, 147901 (2003);
C. W. J. Beenakker and M. Kindermann, Phys.
Rev. Lett {\bf 92}, 056801 (2004);
P. Samuelsson, E. V. Sukhorukov and M. Buttiker,
Phys. Rev. Lett {\bf 91}, 157002 (2003); P. Samuelsson, E.V.
Sukhorukov, M. B\"uttiker Phys. Rev. Lett. {\bf 92}, 026805 (2004); 
A. Di Lorenzo and Yu. V. Nazarov, Phys. Rev. Lett. {\bf 94}, 210601 (2005).
%
\bibitem{loss_divincenzo} D. Loss and D. P. DiVincenzo, Phys. Rev. A
{\bf 57}, 120 (1998).
%
\bibitem{coherence_bulk} J. M. Kikkawa and D. D. Awschalom, Phys.
Rev. Lett. {\bf 80}, 4313 (1998).
%
\bibitem{coherence_dot} J. M. Elzerman, R. Hanson, L. H. W. van
Beveren, B. Witkamp, L. M. K. Vandersypen and
L. P. Kouwenhoven, Nature {\bf 430}, 431 (2004).
%
\bibitem{lesovik_martin_blatter} G. B. Lesovik, T. Martin, and G.
Blatter, Eur.\ Phys.\ J.\ B {\bf 24}, 287 (2001).
%
\bibitem{loss} P. Recher, E. V. Sukhorukov, and D. Loss, Phys.
Rev. B {\bf 63}, 165314 (2001).
%
\bibitem{saraga} W. D. Oliver, F. Yamaguchi, and Y. Yamamoto, Phys.
Rev. Lett. {\bf 88}, 037901 (2002). D. S. Saraga, D. Loss, Phys. Rev.
Lett. {\bf 90}, 166803 (2003); D. S. Saraga, B. I. Altshuler, D. Loss and R. M. Westerwelt, 
Phys. Rev. Lett. {\bf 92}, 246803 (2004). 
%
\bibitem{TP} O. Sauret, D. Feinberg and T. Martin, Eur. Phys. J. B
{\bf 32}, 545 (2003); O. Sauret, D. Feinberg and T. Martin,
Phys. Rev. B {\bf 69}, 035332 (2004).
%
\bibitem{kawabata} S. Kawabata, J. Phys. Soc. Japan {\bf 70}, 1210 (2001); also, full 
counting statistics have been used by F. Taddei and R. Fazio, Phys. Rev. B {\bf 65}, 
075317 (2002); L. Faoro, F. Taddei and R. Fazio, Phys. Rev. B {\bf 69}, 125326 (2004).
%
\bibitem{chtchelkatchev} N. M. Chtchelkatchev, G. Blatter, G. B.
Lesovik and T. Martin, Phys. Rev. B {\bf 66}, 161320(R) (2002).
%
\bibitem{sauret_bloch} O. Sauret, D. Feinberg and T. Martin,
 Phys. Rev. B {\bf 70}, 245313 (2004).
%
\bibitem{these} O. Sauret, PhD Thesis, 
Universit\'e Joseph Fourier, Grenoble (2004, in French).
%
\bibitem{CA} J. M. Byers and M. E. Flatt\'e, Phys. Rev. Lett. {\bf 74}, 
306 (1995); G. Deutscher and D. Feinberg, Appl.\ Phys.\ Lett.\ {\bf
76}, 487 (2000); G. Falci, D. Feinberg, and F. W. J. Hekking,
Europhys.  Lett. {\bf 54}, 255 (2001);
D. Feinberg, Eur. Phys. J.B {\bf 36},419 (2003).
%
\bibitem{choi_bruder_loss} M. S. Choi, C. Bruder and D. Loss, Phys.
Rev. B {\bf 62}, 13569 (2000).
%
\bibitem{CT} D. V. Averin and Yu.  V. Nazarov, in {\it Single Charge
Tunneling}, H. Grabert and M.H. Devoret eds.
(Plenum, New York 1992).
%
\bibitem{buttiker} P. Samuelsson, E.V. Sukhorukov and M. B\"uttiker
Phys. Rev. B {\bf 70}, 115330 (2004).
%
\bibitem{burkard} G. Burkard, D. Loss, and E. V. Sukhorukov Phys. \
Rev. \ B {\bf61}, R16303-R16306 (2000).
G. Burkard and D. Loss Phys. \ Rev. \ Lett. \ {\bf 91}, 087903 (2003).
%
\bibitem{sauret_feinberg} O. Sauret and D. Feinberg, Phys.\ Rev.\
Lett.\ {\bf 92}, 106601 (2004).
%
\bibitem{korotkov} A. N. Korotkov Phys. \ Rev. \ B {\bf 49}, 10381-10392 (1994);
%
\bibitem{note} The initial conditions for calculating correlations functions
within the density matrix equations
should be specified for the coherences
as well as for the populations. At $t=0$ the latter are taken as usual
to be $1$ for one (arbitrary) state
and $0$ for the others. On the other hand, the coherences are naturally taken to be zero at $t=0$.
%
\bibitem{BB} Ya. M. Blanter and M. B\"uttiker, Physics Rep. {\bf 336}, 1
(2000).
%
\bibitem{positive_correlations} J. Torr\`es and T. Martin, Eur. Phys.
J. B {\bf 12}, 399 (1999).
%
\bibitem{doubling_noise} V. A. Khlus, Zh. \'Eksp. Teor. Fiz.{\bf 93}, 2179 (1987)
[Sov. Phys. JETP {\bf 66}, 1243 (1987)]; X. Jehl, P. Payet-Burin, C. Baraduc,
R. Calemczuk and M. Sanquer, Phys. Rev. Lett. {\bf 83}, 1660 (1999).
%
\bibitem{footnote2} This is not true anymore if the two branches are unbalanced, e.g. $\Gamma_L \neq \Gamma_R$.
%
\bibitem{aspect} A. Aspect, J. Dalibard, and G. Roger, Phys.\ Rev.\
Lett.\ {\bf 49}, 1804 (1982); A. Zeilinger, Rev.  Mod. Phys. {\bf71},
S288 (1999).
%
\bibitem{lebedev} A.V. Lebedev, G.B. Lesovik and G. Blatter, Phys. Rev. B{\bf 71}, 045306 (2005).
%
\bibitem{exptCA} D. Beckmann, H. B. Weber and H. v. L\"ohneysen, Phys. Rev. Lett. {\bf 93}, 197003 (2004);
S. Russo, M. Kroug, T. M. Klapwijk and A. F. Morpurgo, Phys. Rev. Lett. 95, 027002 (2005).
\end{thebibliography}
\end{document}